# Low-Light Image Restoration Based on Retina Model using Neural Networks

Yurui Ming, Yuanyuan Liang

*Abstract*—**We report the possibility of using a simple neural network for effortless restoration of low-light images inspired by the retina model, which mimics the neurophysiological principles and dynamics of various types of optical neurons. The proposed neural network model saves the cost of computational overhead in contrast with traditional signal-processing models, and generates results comparable with complicated deep learning models from the subjective perceptual perspective. This work shows that to directly simulate the functionalities of retinal neurons using neural networks not only avoids the manually seeking for the optimal parameters, but also paves the way to build corresponding artificial versions for certain neurobiological organizations.**

## I. INTRODUCTION

SCHOLARS already discover the intricacy of the information processing capabilities of the mammalian's retina [1]. It does intensive pre-processing to the optical signals before forwarding them to higher-level visual cortices. Recent study also reveals the bunch of subtypes of retinal neurons, all of which are endowed with special roles to enable the hosts to swiftly and smartly adjust to varying environments based on perceptions [2, 3]. Although to disclose all the retina's secrecies folded during the evolution is still a long way to go, by resorting to the working principles of retina, more intelligent algorithms related to perception can be designed to generate elegant achievements. Examples along this thread include the retinex theory [4], photoreceptor adaptation proposition [5], etc., which are specifically used to cater to image processing tasks such as high-dynamic range (HDR) image tone mapping (TM) [6-9].

The work in [10] is iconic in that it intensively considers the working principles of various types of neurons, such as horizontal neurons, bipolar neurons, etc., to inspire a model for TM task. Usually, image occurred in nature can be within a rather high dynamic range, which needs to be mapped or compressed in a suitable range that suits the eye. Since human eyes do the TM in some unconscious way, it is believed that the characteristic of retina must be endowed with the capabilities in processing dynamic scenes. In [10], the authors systematically inspect different aspects of the retinal circuitry from the signal processing perspective, ranging from feedbacks of the horizontal cells to activation patterns of the bipolar cells. These neuronal processing traits inspire the authors articulating a corresponding computational model. By separating image into individual channels to modulate by the algorithm and aggregating the outcomes, this work achieves the best result among all.

Usually, a traditional treatment of TM in digital image processing is histogram equalization, namely, to re-adjust the distribution of histogram in an equilibrium conforming the new pixel value range. However, the reshuffle of pixel values tends to cause the original image loses the colour constancy, which means objects can have quite different or unnatural colours after restoration. This is especially challenging for the low-light image restoration (LIIR) task. Although compared with traditional approaches, work in [10] is with exceptional result, however, it is modelling the procedures by mathematical formulas and still too algorithmic-oriented. It uses convolutions and difference of Gaussian (DoG) functions to depict the characteristic of optical signal processing of the corresponding neurons, which potentially loses the biological intuition because of the minimal chance that retina works in such a way. Meantime, some parameters of the model depend on domain knowledge, and it requires experience from the author to select the most appropriate ones.

In this report, inspired by the work in [10], we re-examine the working principles of retina and design a network to tackle the LIIR problem. The network conforms with the processing flow of the optical signal by different neurons and has a clear correspondence between the neural pathway in the retina, in this way it bears a manifest explanation towards the design motivation. Meantime, this simple model merits the end-to-end learning philosophy to avoid manually seeking the optimal parameters. The experiment shows a satisfying image restoration from the subjective perceptual perspective, and we plan the improvement over object metric as future work.

## II. COMPUTATIONAL MODEL

It is already known that for eyes, the cone photoreceptors are dedicated to colours and the rod photoreceptors are dedicated to illuminance. Although in low illumination environment, activations of rod cells over cone cells render a monochromatic or grey image, the downstream cells such as horizontal cells (HCs), amacrine cells (ACs) are still believed to strike for a polychromatic visual perception to benefit the survival [11, 12]. As a consequence, [10] suggests a model which takes two stereotypical pathways for optical signal processing in the retina. The first is the vertical path where signals are picked up by photoreceptors, relayed by bipolar cells (BCs) and sent to ganglion cells (GCs). The second is the lateral pathways where local feedbacks carry information from horizontal cells back to photoreceptors and from amacrine cells to horizontal cells.

In addition, cone photoreceptors mainly consist of three types, namely, S-cones, M-cones and L-cones, each of which is with a different pigment and sensitive to different wavelengths. From the image processing prospective, these cones grossly correspond to the red, green and blue (RGB) channels of a colour image, so in [10], an overall split-then-combine computational flowchart is suggested.

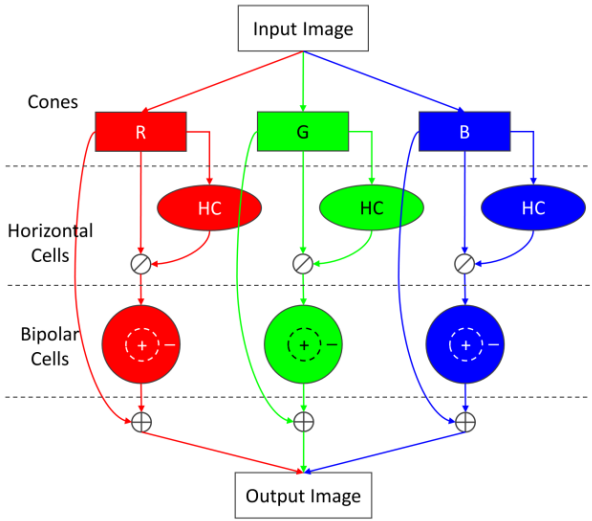

Fig. 1. This is a sample of a figure caption.

However, we take one step further to consider the electrophysiological fact of the neurons in the retina. Usually, signal conduction of neurons is accomplished via action potentials through intermediate neurons in a sequential way, this is the case for most neurons in cerebrum cortices. But the dimensions of some neurons in the retina is so miniature that allows the local graded potentials propagate from the up-stream synapses to the down-stream somas. An example is the bipolar cell, which excites the following ganglion cells directly via local graded potentials from the synapses between photoreceptors and bipolar cells. Therefore, we assume that photoreceptors also exert a direct impact on ganglion cells. Put all these together, we propose a flow of optical signal processing as in Fig. 1.

According to [10], information processed by HCs can be represented in (1), where $I_c(x,y)$ denotes the value of channel $c$ at location $(x,y)$, $g(x,y; \sigma(x,y))$ denotes the kernel at location $(x,y)$ and also depends on the channel value at the same location.

$$h(x,y) = I_c(x,y) * g(x,y; \sigma(x,y)) \qquad (1)$$

However, the circuits prior to BC also incur a recursive modulation to the channel data to produce $b(x,y)$, input value for BCs, as in (2).

$$b(x,y) = \frac{I_c(x,y)}{\alpha + h(x,y)} \qquad (2)$$

Direct implementation of (2) might introduce the numerical instability. Actually, observation of (2) can easily corelate it with the infinite impulse response (IIR) filter from the signal processing perspective. Since to some extent, IIR can be approximately implemented as a FIR (finite impulse response) filter, therefore, we simplify (2) to a more straightforward FIR form as in (3).

$$b(x,y) = \alpha \cdot I_c(x,y) + \beta \cdot I_c(x,y)h(x,y) \qquad (3)$$

If we substitute the definition of $h(x,y)$ in (3) with (1), the overall form roughly resembles the residual structure. Considering the exact role played by HC is still under research, we take a further step to re-write (3) just in a residual form as (4).

$$b(x,y) = I_c(x,y) + h(x,y) \qquad (4)$$

Next, along the optical signal processing neural circuitry, the BCs exert a double-opponent effect on the signal modulated by HCs, which can be modelled as a convolution with a DoG kernel $f$, as in (5):

$$v(x,y) = b(x,y) * f(x,y) \qquad (5)$$

Generally speaking, the neural network lacks the constraint to exert such an effect, namely, the learned filters embedded with DoG property. Therefore, we only require the initialization of corresponding convolutional layer weights to comply with a DoG kernel instance. For this purpose, we choose $\sigma_1 = 0.5$, $\sigma_2 = 5$, and construct the filter according to (6). The final sampled values of size $5 \times 5$ are shown in (7):

$$k = G(0, \sigma_1) - G(0, \sigma_2) \qquad (6)$$

$$k = \begin{bmatrix} -0.0369 & -0.0391 & -0.0397 & -0.0391 & -0.0369 \\ -0.0391 & -0.0303 & 0.0413 & -0.0303 & -0.0391 \\ -0.0397 & 0.0413 & 0.5754 & 0.0413 & -0.0397 \\ -0.0391 & -0.0303 & 0.0413 & -0.0303 & -0.0391 \\ -0.0369 & -0.0391 & -0.0397 & -0.0391 & -0.0369 \end{bmatrix} \quad (7)$$

Further, as we mentioned above, bipolar cells relay signals not via action potential but via local potential, so we assume the signals which are attenuated but still potentially exert influence on the ganglion cells. The assumption refines (5) to (8) in below:

$$v(x,y) = I_c(x,y) + b(x,y) * f(x,y) \qquad (8)$$

Based on the above computational model, we architect the neural network in the next section.

## III. EXPERIMENT AND RESULT

### A. Dataset

We use the open-sourced LOw-Light (LOL) image dataset to conduct our experiment [13]. This dataset is composed of 500 low-light and normal-light image pairs and divided into 485 training pairs and 15 testing pairs. All the images have a resolution of 400×600 with most of them covering indoor scenes. Since the image size is similar to that of most images captured with applications such as surveillance during low-light conditions, we directly process the images without down-sampling to reflect the real cases.

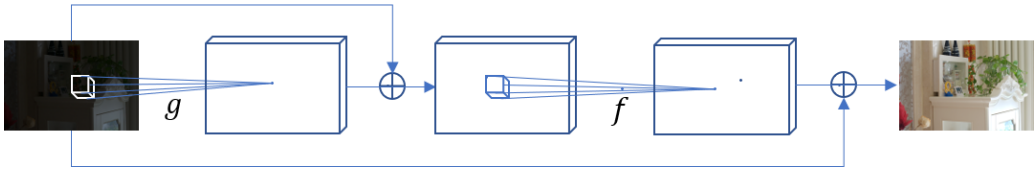

Fig. 2. Architecture of the neural networks used in this work. Two vanilla depthwise convolutional networks in residual form.

TABLE I
CONFIGURATION OF THE NETWORK STRUCTURE

| Op. Name | Op. Type | Multiplier | #Filters | Size | Stride | Act. | Pad. |
|---|---|---|---|---|---|---|---|
| g | DepthwiseConv2D | 1 | 1 | (3, 3) | 1 | ReLU | Same |
| f | DepthwiseConv2D | 1 | 1 | (5, 5) | 1 | ReLU | Same |

### B. Network Structure and Configuration

From Fig. 1, it is manifest that channel values of RGB are processed respectively, therefore, depthwise convolutions are used here to maintain the separation [14]. Fig. 2 demonstrate the overall network structure in accordance with (1), (4) and (8), and TABLE I details the configuration of each operation. Notably, in (1) the $\sigma$ parameter of convolution $g$ depends on the pixel values when the operation slides over the monochromatic input $I_c(x, y)$. However, we eliminate the process of determining $\sigma$ for each location to avoid the computation overhead, which means we only specify the configuration to let the network learn the optimal weights during training.

With the above configuration, we train the network for 20 epochs with learning rate of 0.001 and batch size 8. Since there are only 108 learnable parameters, the possibility of overfitting is too low to carry out extra validation. Therefore, the network is directly running against the test set after training. Fig. 3 shows results of two test samples. It is manifest the simple network is capable of restoring the low-light image as to certain perceptive satisfactoriness. However, as in TABLE I, the limited size of the filters may not be able to extract the global illumination information and uphold it to some decent level, which results in overall darkness of the restored images compared with the ground-truth ones.

To objectively assess the quality of restored image, the structural similarity index measure (SSIM) is calculated against all test samples to indicate the overall performance. The average SSIM is 36.2%, which is lower than the reported SSIM ranging from 76.3% to 93.0% achieved by other various models. However, previous research which relies on heavy models are more theoretic-oriented, our super simple network can guarantee the potential deployment in different systems and scenarios.

### IV. CONCLUSION

We designed a simple neural network inspired by the working principles of various neurons in the retina and applied it to the LIIR problem. By experimenting on the LOL dataset which is used to benchmark the LIIR task, our results show certain satisfactoriness of restored images from the subjective perception perspective. Although the objective assessment from SSIM lags behind other methods, we believe inspiration from biological computation to construct simple networks matters and further work can improve the current limitation.

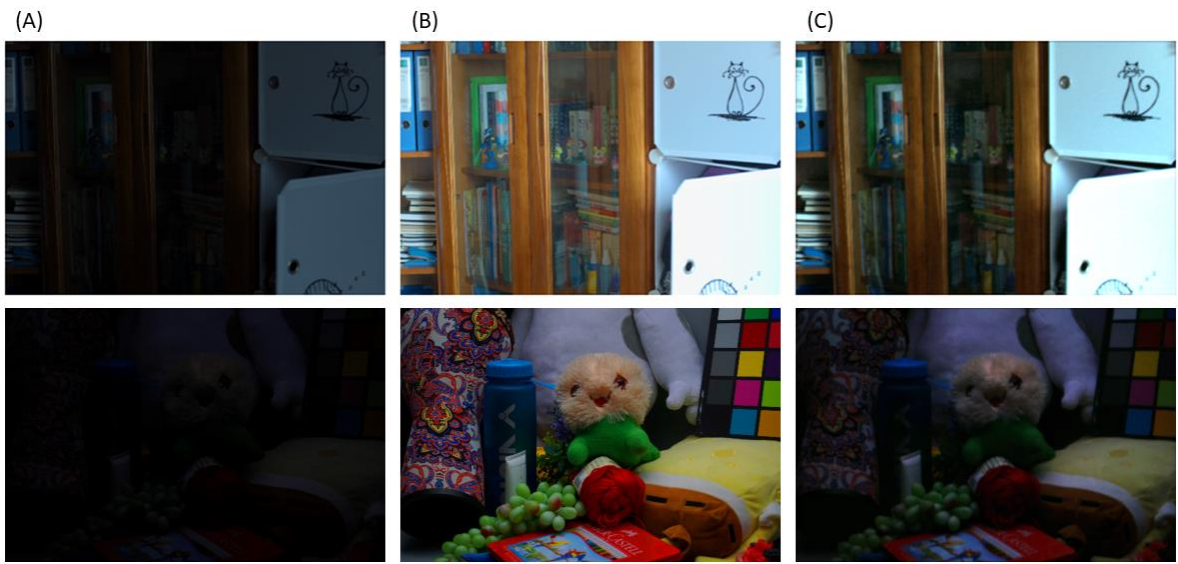

Fig. 3. (A) Images under low-light condition; (2) Ground-truth Images. (3) Restored images by the trained network.